\documentclass[12pt,a4paper]{article}
\usepackage[a4paper,margin=1.3in]{geometry}
\usepackage{parskip}
\usepackage[numbers]{natbib}
\usepackage{graphicx}
\usepackage[hidelinks]{hyperref}
\usepackage{float}

\title{A Scan-Based Analysis of Internet-Exposed IoT Devices Using Shodan Data}

\author{
Richelle Williams\thanks{Email: \href{mailto:williamsr2023fau.edu}{williamsr2023fau.edu}}
\and
Fernando Koch\thanks{Email: \href{mailto:kochf@fau.edu}{kochf@fau.edu}} \\
Department of Electrical Engineering and Computer Science \\
Florida Atlantic University, Boca Raton, FL, USA
}

\date{} 

\begin{document}
\maketitle
\abstract{


An open measurement problem in IoT security is whether scan-observable network configurations encode population-level exposure risk beyond individual devices. We analyze internet-exposed IoT endpoints using a controlled multi-country sample from Shodan Search and Shodan InternetDB, selecting 100 hosts identified via TCP port 7547 (TR-069/CWMP) and evenly distributed across the ten most represented countries. Hosts are enriched with scan-derived metadata and analyzed using feature relevance assessment, cross-country comparison of open and risky port exposure, and supervised classification of higher-risk exposure profiles. The analysis reveals consistent cross-country differences in exposure structure, with mean risky-port counts between 0.4 and 1.0 per host, and achieves a balanced accuracy of approximately 0.61 when classifying higher-risk exposure profiles.

}


\section{Introduction}
\label{sec:intro}

Internet-reachable configuration choices influence exposure risk in large-scale Internet of Things (IoT) deployments \cite{LeeLee2015IoT}. Devices that expose management or auxiliary services to the public Internet permit unsolicited interaction and automated probing, creating an externally observable attack surface independent of documented device vulnerabilities \cite{abdulqawy2015iot}. Such configurations accumulate across deployments, producing population-level exposure that can be quantified using scan-derived data across heterogeneous networks and administrative domains.

Internet-wide scanning enables large-scale measurement of externally reachable systems by characterizing exposed services using protocol responses, service banners, and metadata from public-facing endpoints, supporting comparative analysis across geographic regions. Remote management protocols such as TR-069 (CWMP), commonly deployed for customer premises equipment, constitute a significant portion of the Internet-visible IoT service surface, and hosts exposing TCP port 7547 represent a representative population of publicly reachable managed embedded systems.

\begin{figure}
    \centering
    \includegraphics[width=1.2\linewidth]{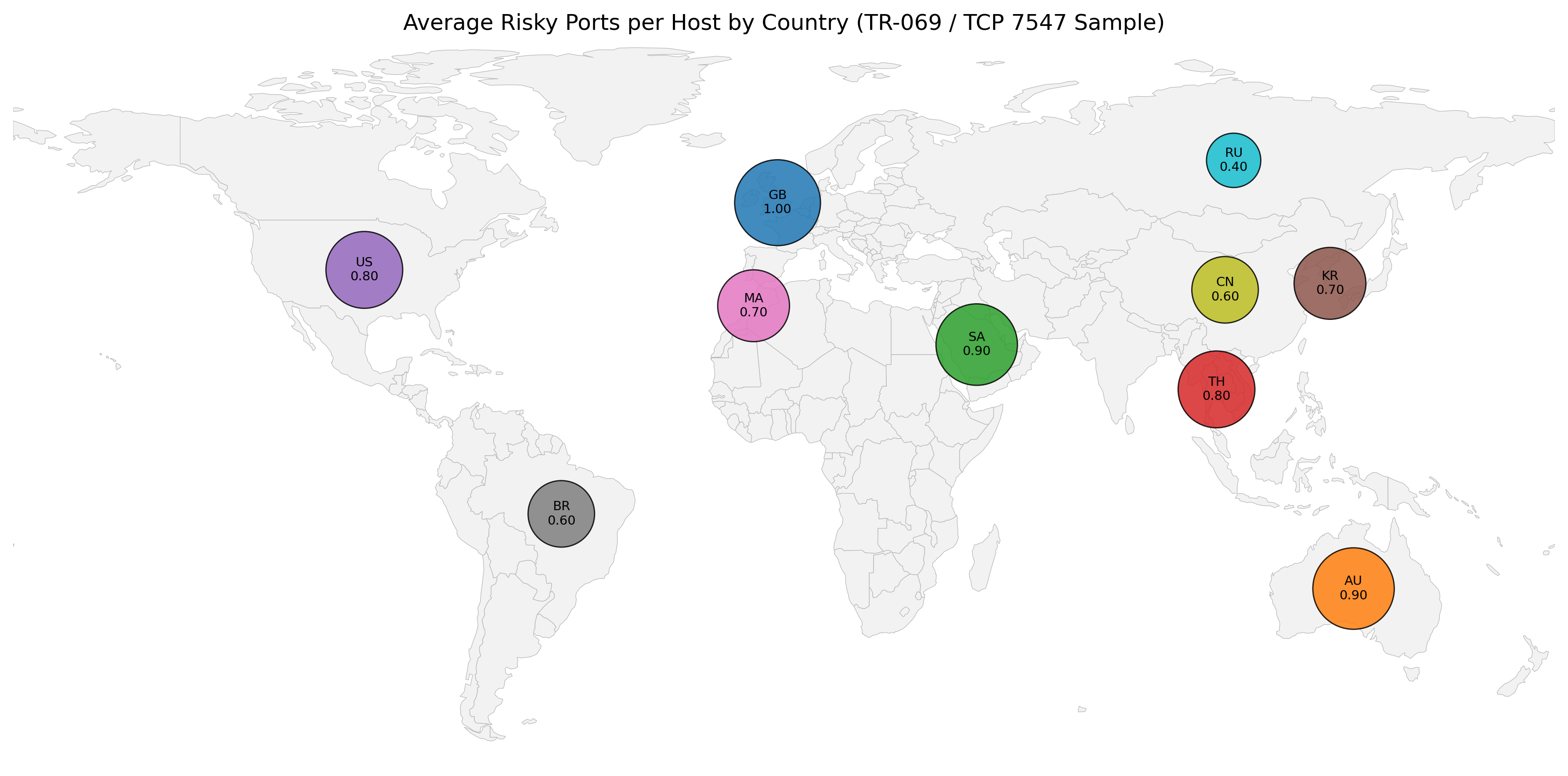}
    \caption{Average risky ports per host for TR-069–exposed devices (TCP port 7547) by country under fixed sampling conditions. Circle size reflects the mean risky-port count per host.}
    \label{fig:consolidated_results}
\end{figure}

Figure~\ref{fig:consolidated_results} shows consistent variation in exposure indicators across geographic groups within the TR-069–exposed population, indicating that externally observable service configuration characteristics differ systematically under fixed sampling conditions. Using Shodan InternetDB data for internet-reachable hosts exposing TCP port 7547 (TR-069/CWMP), the analysis examines open-port counts, service surface area, and co-exposed services across controlled multi-country samples, demonstrating that scan-derived attributes enable population-level differentiation of higher- and lower-risk exposure profiles. In contrast to prior IoT security research, which primarily focuses on vulnerability discovery, exploit development, and device-level compromise \cite{7342646} and establishes that exposed services facilitate attack, this analysis provides evidence of a reproducible population-level relationship between externally observable configuration characteristics and exposure risk using Internet-wide scanning and controlled geographic comparison \cite{vanKranenburgBassi2012IoTChallenges}.

This study addresses that measurement gap by evaluating whether scan-derived service configuration attributes can be used to quantify and compare IoT exposure risk under controlled sampling conditions using Internet-wide scanning data and compare IoT exposure risk across geographic groups under controlled sampling conditions, demonstrating the feasibility of population-level exposure assessment using externally observable data without reliance on exploit execution, device access, or vendor cooperation. Through a series of Internet-scale analyses, the study establishes the analytical value of this approach for characterizing exposure risk based solely on scan-derived configuration attributes.

\begin{itemize}
    \item a population-level measurement of IoT exposure based on scan-observable service configuration characteristics derived from Internet-wide data;
    
    \item a controlled multi-country analysis of service surface area and exposure indicators for internet-reachable hosts exposing TR-069 (TCP port 7547)
    
    \item an empirical evaluation of the predictive value of scan-derived host attributes for identifying higher-risk exposure profiles.
\end{itemize}

\section{Background}
\label{sec:background}

The proliferation of Internet-connected embedded systems has resulted in a growing population of IoT devices reachable from public network address space. Constrained computational resources, inconsistent patch management practices, and extended operational lifecycles contribute to elevated security and privacy exposure across consumer and enterprise IoT deployments \cite{Wang2021IoTEvolution}. Devices deployed in smart home environments, organizational infrastructures, and industrial monitoring systems are exposed through remote management services and vendor-specific administrative interfaces with permissive access controls.

Internet-wide scanning enables measurement of externally reachable systems at scale. Tools such as ZMap support discovery of exposed services and large-scale security measurement across the global address space \cite{Balaji2019IoTSurvey}. Scan-derived datasets support indexing and querying of exposed hosts using service banners, protocol fingerprints, and metadata retrieved from public-facing endpoints \cite{LuXu2019IoTCybersecurity}. These approaches enable comparative analysis of exposure patterns across geographic regions, network operators, and device populations.

Public vulnerability documentation supports structured analysis of software and firmware weaknesses. CVE-indexed disclosures and NVD-curated records provide standardized identifiers and severity metrics for systematic vulnerability assessment\cite{article}. These repositories enable association between observed device attributes and documented vulnerability conditions using standardized identifiers and are commonly used with scan-derived measurements.

Prior research uses scan-observable evidence to categorize IoT devices and assess exposure. Classification techniques leverage protocol behavior, banner content, and service metadata to infer device type and deployment characteristics. Internet-scale studies report widespread insecure service exposure and persistent misconfigurations across diverse IoT populations \cite{Zikria2021NextGenIoT}. Longitudinal analyses report stability in exposure prevalence across common device categories and deployment contexts \cite{Freescale2014IoT}.

Empirical evidence of exploitation highlights risks associated with exposed IoT endpoints. The Mirai botnet compromised Internet-accessible devices through weak authentication and outdated configurations, enabling coordinated attacks at scale \cite{Lee2017IoTSurvey}. Subsequent analyses link exposed IoT endpoints to distributed denial-of-service campaigns and sustained network disruptions \cite{Suresh2014IoTReview}, demonstrating an association between externally observable exposure and exploitation activity.

Many large-scale assessments model IoT systems as generic exposed hosts, limiting attribution to specific device classes or product families. Scan-derived identifiers often contain incomplete or noisy attributes due to banner suppression, network middleboxes, reverse proxies, and vendor-specific protocol implementations. These factors reduce classification precision and complicate reliable mapping to firmware versions and vulnerability conditions.

Many studies associate vulnerabilities through direct matching of observed banners or version strings to public vulnerability records. This approach is constrained by missing or ambiguous version metadata and may produce incorrect associations when vulnerability conditions depend on configuration properties not observable through external scans. Embedded firmware analyses report recurring security weaknesses and limited update coverage across device ecosystems \cite{Perwej2019IoTDomains}, further complicating accurate exposure assessment.

Internet-scale IoT security measurement requires methods that integrate geographic exposure characterization, device-type attribution, and vulnerability association under consistent analytical assumptions. Scanning infrastructure supports large-scale observability \cite{Chernyshev2018IoTSimulatorsTestbeds}, while exposure studies quantify risk-relevant weaknesses using scan-derived evidence \cite{Lee2020IoTCybersecurity}. Incomplete metadata continues to limit device categorization accuracy and vulnerability attribution.

Feature-based modeling has been applied as an alternative approach for vulnerability inference when direct banner-to-vulnerability mapping is unreliable. Prior work applies machine learning techniques to IoT identification and behavior characterization using network-observable features, including exposed ports, protocol responses, manufacturer strings, and geographic attributes\cite{Lone2023IoTCybersecurity}. This study applies scan-derived features to perform country-level exposure characterization, device-category security posture analysis, and supervised evaluation of vulnerability association using externally observable host attributes.

\section{Method}
\label{sec:method}
This study measures exposure structure in internet-reachable IoT systems to evaluate whether scan-observable attributes reflect security posture. The workflow follows a reproducible design with transparent data collection and bias-aware sampling consistent with Internet measurement practices \cite{Dritsas2025IoTSurvey}. Internet-wide scanning is used to observe exposed services at scale \cite{CHOO2021102136}, and the analysis combines descriptive measurement, cross-country comparison, and supervised classification to support population-level assessment. The data consist of scan-derived, observational network measurements describing externally observable service configurations, including exposed ports, services, and associated metadata, captured as a cross-sectional snapshot at the host level without authenticated access, exploit execution, or internal configuration visibility.

Internet-reachable hosts are identified using Shodan Search and enriched using Shodan InternetDB, which provides open ports, tags, CPE identifiers, and vulnerability-related metadata when available \cite{Alferidah2020IoTBigData}. Only processed, derived data are retained in a structured tabular format to support repeatable analysis. The seed dataset is constructed using a Shodan query targeting TCP port 7547, which is commonly associated with CWMP (TR-069) on customer-premises equipment \cite{Sam2022IoTWearableSurvey}. Ten countries with the highest host counts are selected, and ten unique hosts are sampled per country, yielding a fixed-size dataset of 100 hosts for controlled cross-country comparison.

Each record includes the following fields:
\begin{itemize}
    \item \textit{group}: integer country-group identifier (1--10).
    \item \textit{country}: ISO country code.
    \item \textit{ip}: public IPv4 address.
    \item \textit{source\_query}: Shodan query string formatted as \texttt{port:7547 country:"XX"}.
\end{itemize}

The dataset contains the following country groups: United States (US), China (CN), South Korea (KR), Brazil (BR), Russia (RU), United Kingdom (GB), Saudi Arabia (SA), Morocco (MA), Australia (AU), and Thailand (TH). The fixed sampling design enables cross-country comparison under uniform sample size constraints.

The threshold of 0.8 is chosen to represent the upper range of normalized mean risky exposure values observed in the sample. This cutoff distinguishes country groups with consistently higher exposure from those with moderate or lower exposure under uniform sampling conditions.

\subsection{Feature Relevance Analysis}

The feature relevance analysis is included to identify which externally observable host attributes are most closely associated with exposure indicators in the dataset. This experiment is selected to examine the contribution of individual scan-derived features such as service counts and metadata presence to exposure differentiation, independent of host prevalence. By comparing host-level attributes to the vulnerability association label using descriptive ranking and correlation-based methods, the analysis determines whether measurable exposure signal is present when fingerprint and metadata coverage is incomplete.

The country-level exposure profile comparison extends this analysis by examining how aggregated exposure indicators vary across geographic groups under fixed sample sizes. This experiment is chosen to assess whether observed differences in exposure arise from variation in service composition rather than uniform reachability. By analyzing mean open-port counts, risky service counts, and metadata availability rates across country groups, the analysis characterizes structural differences in exposure profiles and demonstrates heterogeneous exposure patterns based on externally observable service configurations.

The feature relevance analysis identifies scan-observable host attributes associated with exposure indicators in the dataset. First,  host-level attributes are extracted from Shodan InternetDB enrichment. The feature set includes open port lists, scan tags, CPE identifiers, and vulnerability metadata fields when present. The workflow constructs exposure indicators that include total open ports per host, risky-port exposure counts, and product fingerprint indicators derived from CPE fields.The workflow evaluates feature relevance using descriptive ranking and correlation-based comparisons between exposure indicators and the vulnerability association label.

\begin{figure}[H]
    \centering
    \includegraphics[width=1.0\linewidth]{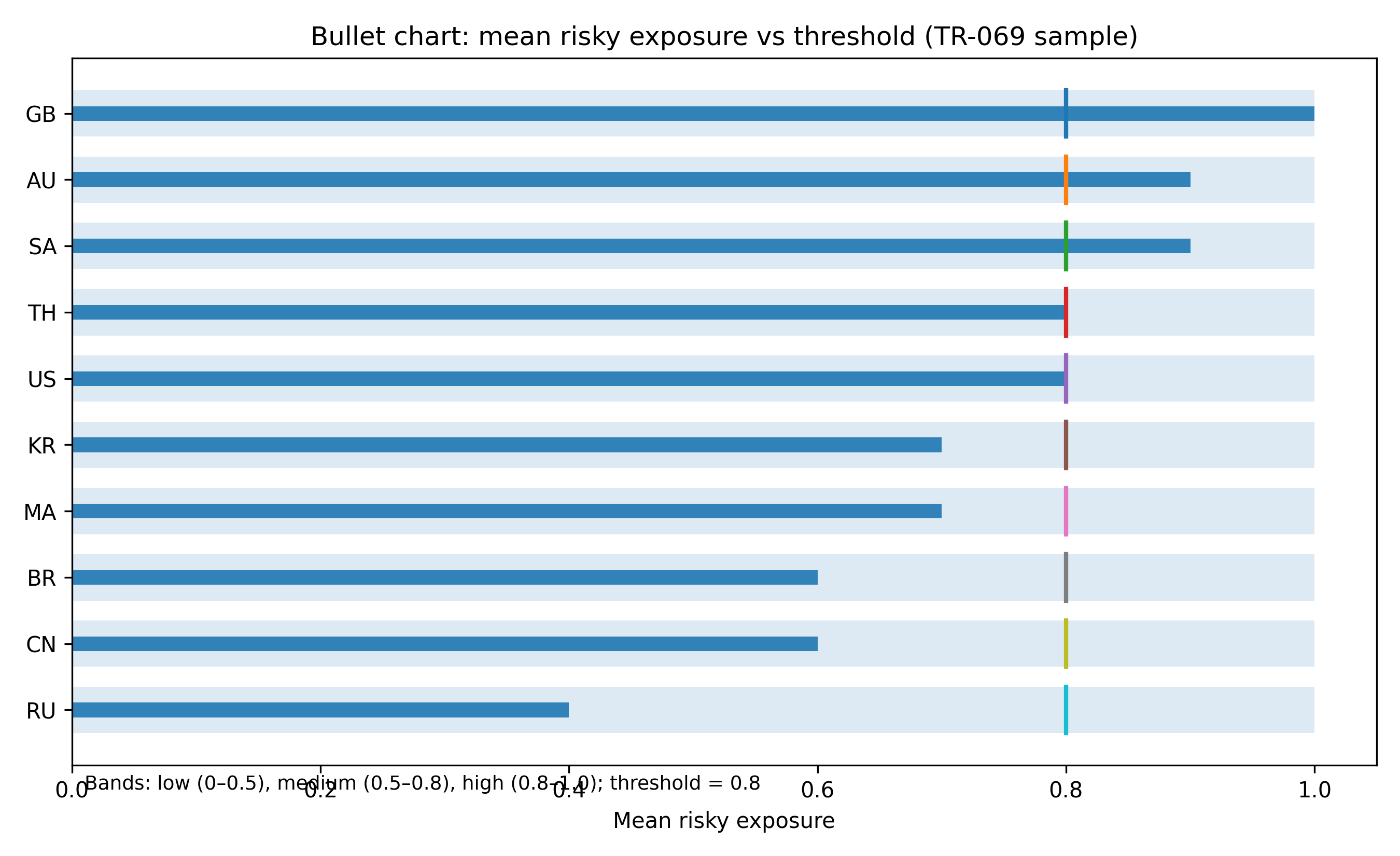}
    \caption{Threshold-based comparison of mean risky service exposure across country groups for TR-069–exposed hosts.}
    \label{fig:bullet_risky}
\end{figure}

Exploratory analysis was performed to assess country-level variation in mean risky exposure using a threshold-based representation. The bullet chart encodes the mean risky service exposure per host for each country group, measured as the average count of externally observable risky services normalized to a fixed scale. Qualitative exposure bands are defined as low (0–0.5), medium (0.5–0.8), and high (0.8–1.0) to contextualize relative exposure levels.  This visualization enables direct comparison of country-level exposure metrics against a consistent benchmark under uniform sampling conditions, highlighting differences in exposure magnitude derived from scan-observable service configurations.

Hosts with higher open-port counts show higher vulnerability association rates than hosts with fewer open ports. Hosts with identifiable product metadata show higher vulnerability association rates than hosts without CPE indicators.

Scan-derived service configuration indicators provide measurable signal for differentiating exposure profiles under incomplete metadata conditions.

\subsection{Country-Level Exposure Profile Comparison}

The country-level exposure profile comparison is conducted to examine structural differences in exposure patterns across geographic groups under fixed sample size conditions. This experiment is selected to move beyond host-level observations and assess whether aggregation reveals systematic variation in exposure structure at the population level. By computing group-level summaries of open-port counts, risky service counts, and metadata availability rates, the analysis provides a consistent basis for comparing exposure characteristics across countries using externally observable scan data.

This experiment is chosen to determine whether exposure variation reflects differences in service surface composition rather than differences in host presence alone. Distributional and rank-based analyses are applied to aggregated indicators to characterize heterogeneity in exposure profiles across country groups. The resulting measurements show that higher-exposure groups exhibit larger service surface areas, as reflected in elevated risky service counts and auxiliary service co-exposure, while lower-exposure groups show more constrained configurations, indicating that exposure differences are driven primarily by scan-observable service characteristics.

 measurement differences are analyzed in exposure structure across country groups under fixed sample size conditions. Host-level exposure indicators by country group are aggregated. The workflow computes mean open-port counts, mean risky-port counts, and metadata availability rates within each group. 

\begin{figure}[H]
    \centering
    \includegraphics[width=0.5\linewidth]{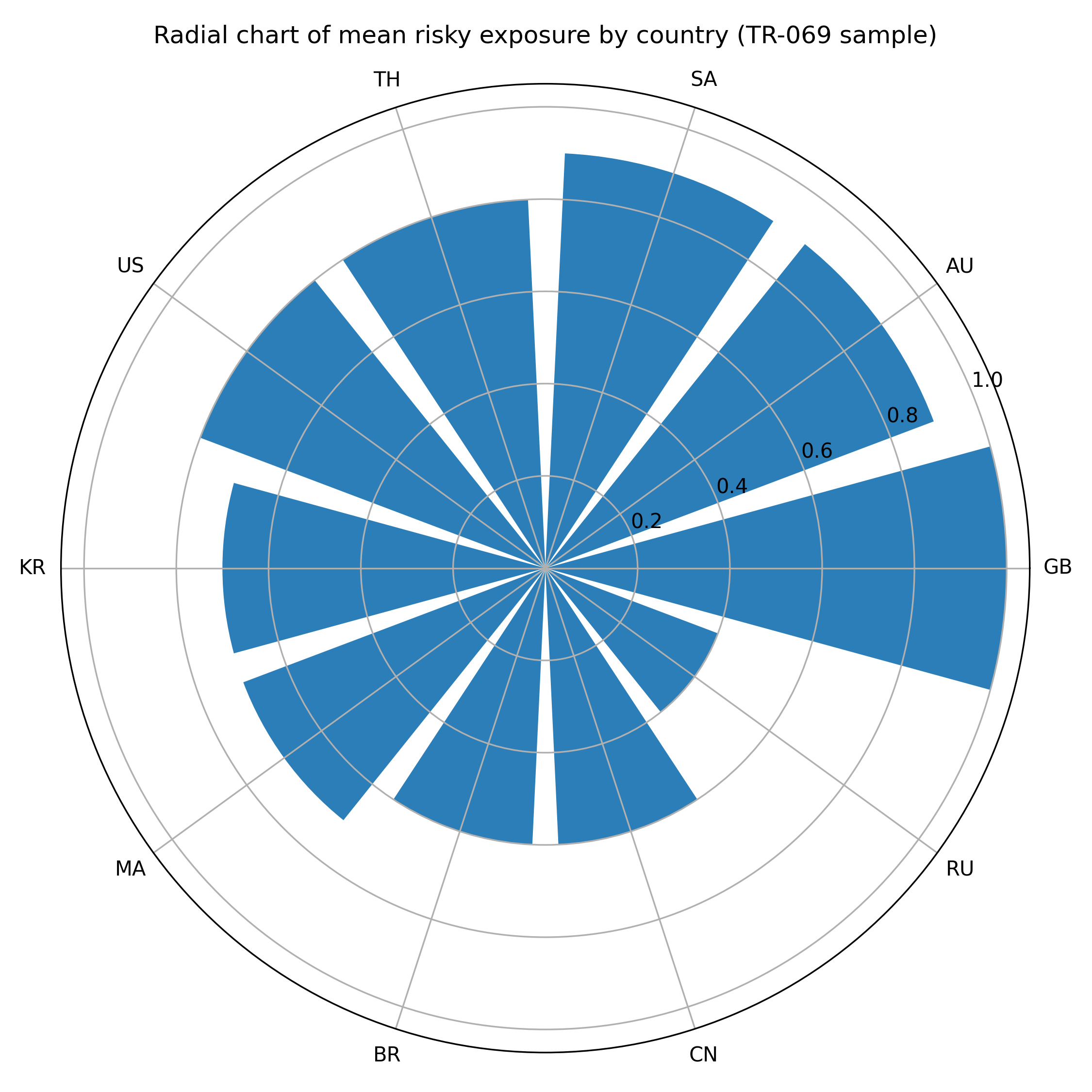}
    \caption{Radial visualization of mean risky service counts per host across country groups in the TR-069 dataset.}
    \label{fig:radial_risky}
\end{figure}

Distributional analysis shows that the data was conducted to characterize variation in scan-derived exposure indicators and to assess the relative impact of host-level features on exposure differentiation. The analysis examined externally observable attributes, including total open-port count, risky service count, and service co-exposure patterns, using descriptive statistics and rank-based comparisons across country groups.

Figure~\ref{fig:radial_risky} summarizes mean risky service exposure by country using a radial representation. Mean exposure values range from 0.40 to 1.00 across the sampled groups, indicating substantial heterogeneity under fixed sampling conditions. Higher-exposure groups exhibit consistently larger service surface areas, reflected in elevated risky service counts and increased auxiliary service co-exposure.

Feature impact was evaluated by measuring the association between scan-derived attributes and exposure indicators using correlation analysis and relative rank separation. Open-port count and risky service presence demonstrate the strongest association with elevated exposure, while basic reachability indicators show limited differentiation. These observations indicate that exposure variation is primarily driven by service surface characteristics rather than host presence alone.

All exploratory measurements are based exclusively on externally observable scan data and reflect cross-sectional host configurations at the time of observation.

Country groups show differences in mean open-port counts. Country groups show differences in mean risky-port counts. Groups with higher metadata availability show higher rates of vulnerability association flags.

Host-level exposure indicators vary across country groups under uniform sampling.

\subsection{Vulnerability Association Prediction}
This experiment applies vulnerability association prediction as an empirical validation of whether scan-derived exposure indicators encode meaningful risk-related signal. A supervised classification model is constructed using engineered features derived from Internet-wide scans, and the target label is defined using vulnerability association metadata from Shodan InternetDB. Classification performance is assessed on held-out data using accuracy, balanced accuracy, and confusion matrix analysis, enabling evaluation of prediction behavior across exposure classes under class imbalance based solely on externally observable attributes.

The results are contextualized through an aggregate analysis of exposure composition across country groups using normalized scan-derived indicators. The streamgraph visualization captures relative differences in risky service presence versus non-risky remainder under fixed sampling conditions, isolating variation in service composition from differences in host prevalence. Observed patterns show that groups with higher predicted exposure exhibit a greater concentration of risky service indicators, while lower-exposure groups are dominated by non-risky components, reinforcing that exposure differentiation is driven by observable service structure rather than uniform reachability.

Vulnerability association prediction is tested  using scan-derived host attributes. In this experiment a supervised classification model using engineered scan-derived features is trained. Predictors include total open ports, risky-port exposure, scan tags, and CPE-derived product indicators when present. The workflow defines the target label using the vulnerability association metadata field from Shodan InternetDB enrichment. The workflow applies a train-test split using stratified sampling when feasible. The evaluation reports accuracy, balanced accuracy, and confusion matrix outputs.

\begin{figure}[H]
    \centering
    \includegraphics[width=0.75\linewidth]{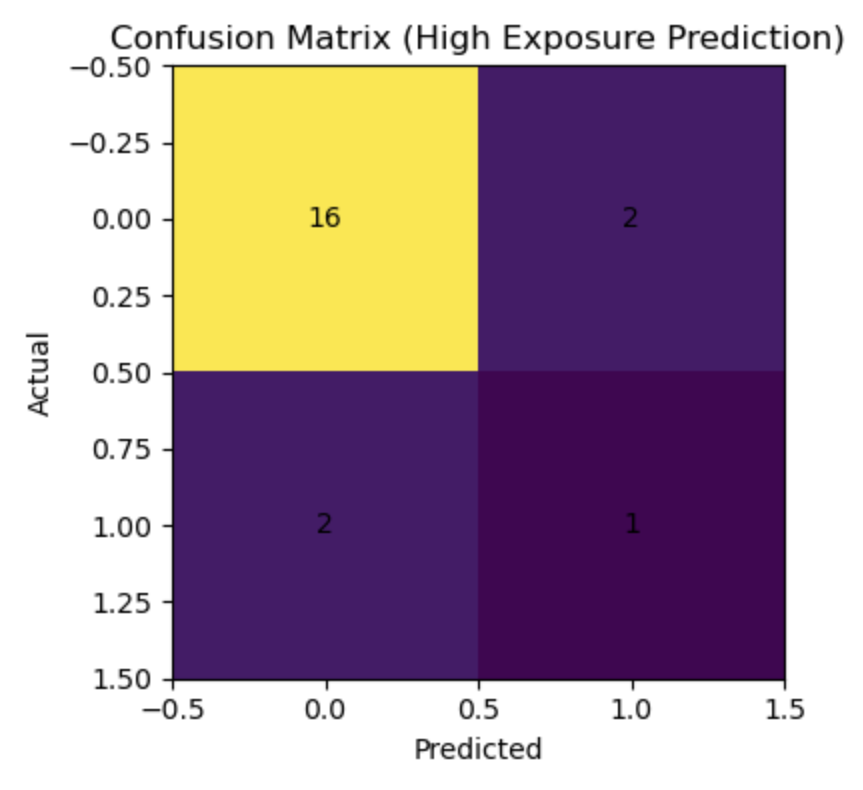}
    \caption{Confusion matrix showing classification performance for predicting high-exposure IoT hosts using scan-derived features.}
    \label{fig:exp3_confusion_matrix}
\end{figure}

The descriptive analysis examines the supervised classification results was conducted using a confusion matrix to examine prediction behavior across exposure classes. The matrix reports true positives, true negatives, false positives, and false negatives for the binary high-exposure label derived from vulnerability association metadata. From these counts, accuracy and balanced accuracy metrics are computed to account for class imbalance within the sampled dataset. The confusion matrix enables assessment of model sensitivity to high-exposure hosts and specificity for lower-exposure hosts, providing insight into error distribution beyond aggregate accuracy measures. All evaluation metrics are computed using held-out test data and reflect classification performance based solely on scan-derived host attributes.

Balanced accuracy exceeds the 0.50 baseline under uniform sampling. Port exposure indicators contribute the strongest predictive signal.

Scan-derived service exposure features support vulnerability association prediction under incomplete fingerprint conditions.

\begin{figure}[H]
    \centering
    \includegraphics[width=1.0\linewidth]{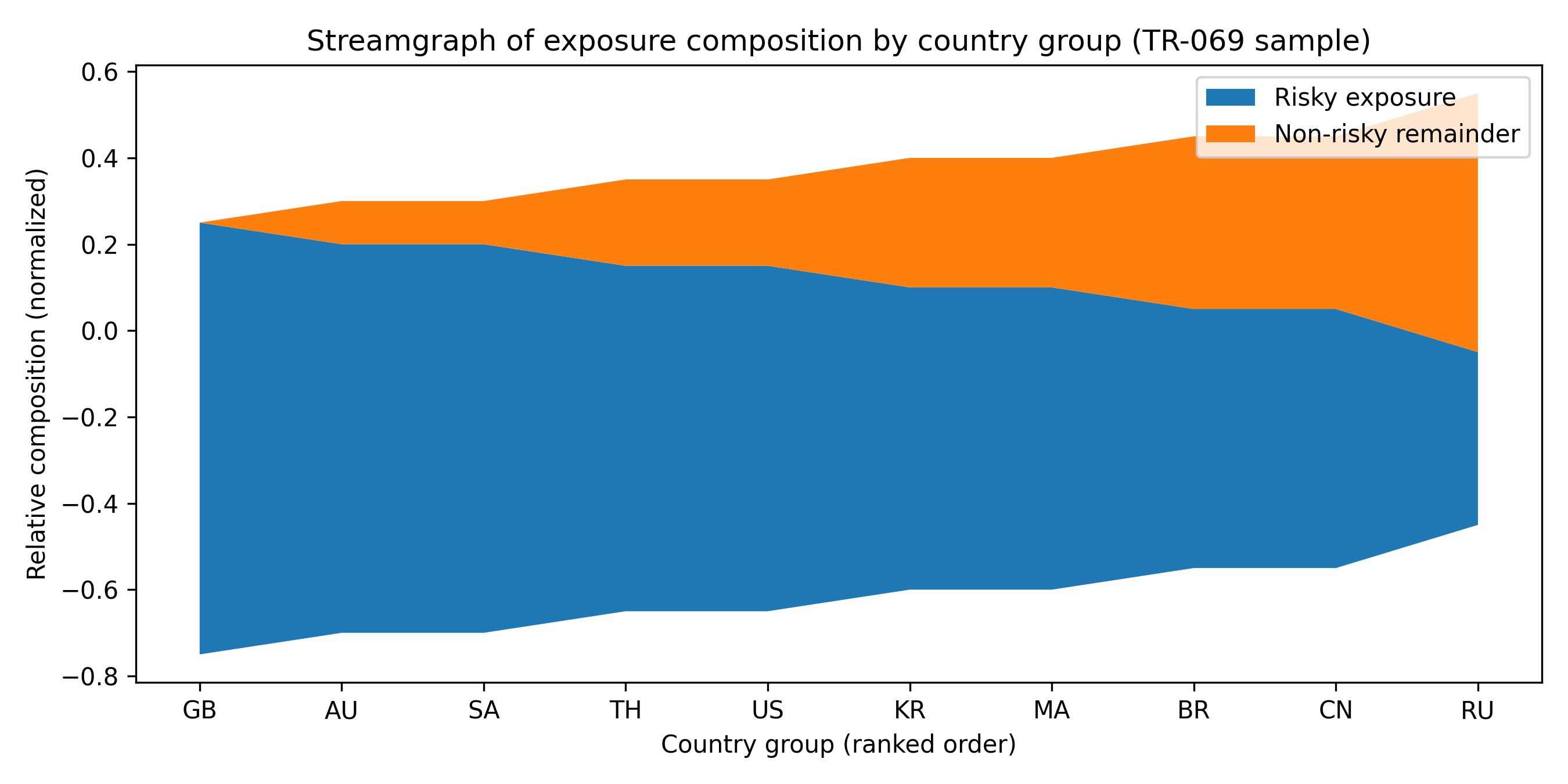}
    \caption{Streamgraph depicting variation in exposure composition across country groups based on scan-derived service indicators.}
    \label{fig:streamgraph}
\end{figure}

An empirical assessment examines the composition of scan-derived exposure indicators across country groups using normalized metrics. The streamgraph encodes the relative contribution of risky service exposure and non-risky remainder for each country group, where values are normalized to represent proportional composition rather than absolute counts. Risky exposure reflects the mean presence of externally observable auxiliary and administrative services per host, while the non-risky remainder represents the complement of this measure within the normalized exposure profile. Country groups are ordered consistently across the horizontal axis to facilitate comparative assessment of exposure composition under fixed sampling conditions. This representation highlights variation in exposure structure across groups while preserving relative differences in service composition derived from externally observable scan data.

The exploratory analysis shows clear variation in exposure composition across country groups. Higher-exposure groups exhibit a larger proportion of risky service presence relative to the normalized exposure profile, while lower-exposure groups show a greater non-risky remainder. This pattern indicates that differences in exposure are driven primarily by service composition rather than uniform host reachability across groups.

Exposure differences are driven primarily by service composition rather than uniform host reachability.

Service exposure indicators show the strongest association with vulnerability-related metadata across experiments. Country grouping partitions the dataset into uniform subsets for comparison. Host-level fingerprints and exposed port combinations separate exposure profiles under fixed sampling conditions. The combined results support scan-derived feature engineering for exposure structure analysis and vulnerability association modeling.

\section{Analysis}
\label{sec:analysis}

The analysis aggregates results across all experiments and maps results to the research questions. The workflow measures exposure indicators, ranks feature relevance, and trains supervised classifiers using scan-derived metadata.

\textit{Host-level service exposure indicators explain exposure variation more effectively than country group membership.}

The Feature Relevance Analysis bullet chart measured associations between scan-observable attributes and exposure indicators in the 100-host Shodan InternetDB sample. The dataset includes ten hosts per country group. Figure~\ref{fig:bullet_risky} reports mean risky-port counts by country group. Hosts with larger exposed port sets show higher risky-port counts. Hosts with richer fingerprint metadata show higher exposure indicator values. Country membership does not separate higher-risk exposure profiles consistently.

Host-level configuration features provide the strongest explanatory signal in the measured sample. Prior scan-based measurement research reports similar behavior for service-level indicators and scan-derived metadata fields \cite{Hwang2015IoTSecurityPrivacy}.

The country-level exposure profile comparison radial chart compared country groups using aggregated host-level exposure metrics. Figure~\ref{fig:radial_risky} reports mean open ports per host by country group. The dataset controls host counts per group. Country groups differ in mean open-port counts and mean risky-port counts under fixed sampling.

Open-port combinations measure service surface breadth. Figure~\ref{fig:streamgraph} ranks predictors for the higher-exposure label. Service-level indicators rank above geographic membership in feature importance.

\textit{Scan-derived metadata supports classification of higher-risk exposure profiles under limited sample size and label imbalance.}

 The vulnerability association prediction confusion matrix and steamgraph trained a supervised classifier using scan-derived host attributes from Shodan InternetDB enrichment. The classifier achieved performance above a random baseline. Figure~\ref{fig:exp3_confusion_matrix} shows correct classification for most low-exposure observations. The classifier misclassifies a larger fraction of high-exposure observations due to minority-class support.

The evaluation uses balanced accuracy to control majority-class inflation under label imbalance \cite{SwessiIdoudi2022IoTSecuritySurvey}. Port exposure indicators provide the strongest predictive signal in the trained model.

Shodan InternetDB exposes host-level fields that include open ports, tags, hostnames, CPE identifiers, and vulnerability-related metadata fields \cite{Zhang2019IoTSecurity}. These fields support supervised classification using scan-derived observations.

\textit{Open-port combinations and fingerprint availability drive exposure structure differences under uniform country sampling.}

Exposed service configuration shows the strongest association with vulnerability-related metadata and exposure indicators in the sampled hosts. Country grouping supports comparative aggregation under fixed sampling. Host-level scan attributes provide stronger separation of higher-risk exposure profiles. Figures~\ref{fig:bullet_risky}--\ref{fig:streamgraph} support scan-derived metadata as a measurement layer for population-level IoT exposure assessment.

The sampled population targets TR-069/CWMP exposure through TCP port 7547. Prior incident and measurement studies associate exposed CWMP endpoints with large-scale scanning and exploitation activity \cite{Zhang2015IoTThreats}. The workflow applies repeatable scan-derived observation procedures and bias-aware sampling practices consistent with internet measurement literature \cite{Varga2017AutomationIoT}.

\section{Conclusions}
\label{sec:conclusions}
This work demonstrates that externally observable, scan-derived service configuration characteristics provide a reliable basis for measuring and comparing IoT exposure risk at the population level across geographic groups. Analysis of Internet-reachable hosts exposing TR-069 shows that open-port counts, risky service presence, and service co-exposure patterns vary systematically by country and remain stable under fixed sampling conditions. These findings indicate that externally visible configuration characteristics capture meaningful variation in IoT service surface area without reliance on exploit execution, device access, or vendor cooperation.

These results are important because they establish Internet-wide scanning as a reproducible and scalable method for comparative IoT exposure assessment across heterogeneous networks and administrative domains. By demonstrating that exposure differences are driven by service surface composition rather than uniform reachability, the study provides a measurement framework that complements vulnerability-centric approaches and supports cross-region risk characterization using publicly observable data. Further investigation should extend this framework to additional management and application protocols, evaluate temporal stability of exposure patterns, and assess how configuration-driven exposure indicators relate to downstream security outcomes across broader IoT populations.

\bibliographystyle{IEEEtran}
\bibliography{references}

\end{document}